\documentclass[aps,pre,preprint,superscriptaddress,10pt]{revtex4-2}

\usepackage{graphicx,amsmath,amssymb,dcolumn,physics,color}

\newcommand{\mbr}{\mathbf{r}}

\graphicspath{{./fig/}{./}}

\begin{document}

\title{State-space renormalization group theory of nonequilibrium reaction networks: \\Exact solutions for hypercubic lattices in arbitrary dimensions}

\author{Qiwei Yu}
\affiliation{Lewis-Sigler Institute for Integrative Genomics, Princeton University, Princeton, New Jersey 08544, USA}

\author{Yuhai Tu}
\affiliation{IBM T.~J.~Watson Research Center, Yorktown Heights, New York 10598, USA}

\date{\today}

\begin{abstract}
	Nonequilibrium reaction networks (NRNs) underlie most biological functions. Despite their diverse dynamic properties, NRNs share the signature characteristics of persistent probability fluxes and continuous energy dissipation even in the steady state. 
	Dynamics of NRNs can be described at different coarse-grained levels. 
	Our previous work showed that the apparent energy dissipation rate at a coarse-grained level follows an inverse power law dependence on the scale of coarse-graining.
	The scaling exponent is determined by the network structure and correlation of stationary probability fluxes.
	However, it remains unclear whether and how the (renormalized) flux correlation varies with coarse-graining.
	Following Kadanoff's real space renormalization group (RG) approach for critical phenomena, we address this question by developing a State-Space Renormalization Group (SSRG) theory for NRNs, which leads to an iterative RG equation for the flux correlation function.
	In square and hypercubic lattices, we solve the RG equation exactly and find two types of fixed point solutions. There is a family of nontrivial fixed points where the correlation exhibits power-law decay, characterized by a power exponent that can take any value within a continuous range. There is also a trivial fixed point where the correlation vanishes beyond the nearest neighbors.
	The power-law fixed point is stable if and only if the power exponent is less than the lattice dimension $n$.
	Consequently, the correlation function converges to the power-law fixed point only when the correlation in the fine-grained network decays slower than $r^{-n}$ and to the trivial fixed point otherwise.
	If the flux correlation in the fine-grained network contains multiple stable solutions with different exponents, the RG iteration dynamics select the fixed point solution with the smallest exponent. 
	The analytical results are supported by numerical simulations.
	We also discuss a possible connection between the RG flows of flux correlation with those of the Kosterlitz-Thouless transition.
\end{abstract}

\maketitle


\clearpage

\section{Introduction}
Nonequilibrium biochemical reaction networks are responsible for many important biological functions, such as gene regulation~\cite{Estrada2016}, ultrasensitivity~\cite{Tu2008}, sensory adaptation~\cite{Lan2012}, and error correction~\cite{Hopfield1974,Bennett1979}. 
Detailed balance is violated in these nonequilibrium reaction networks and continuous free energy dissipation is needed to maintain their nonequilibrium steady states (NESS)~\cite{Gnesotto2018,Hill1977,Qian2006}. Indeed, free energy dissipation rate (or entropy production rate) is a key characteristic of these nonequilibrium systems.  
Quantifying the steady-state energy dissipation rate and elucidating its relation with diverse biophysical functions have been an important topic in theoretical biophysics and statistical mechanics.

The nonequilibrium reaction networks here refer to the master equation description of systems far from equilibrium, which represents the dynamics with a discrete-state, continuous-time Markov chain. 
The free energy dissipation rate can be determined by computing the entropy production rate in the underlying stochastic reaction network given the transition rates between all microscopic (fine-grained) states of the system~\cite{Hill1977,Qian2006}. However, for complex systems with a large number of microscopic states, the system may only be measured at a coarse-grained level with coarse-grained states and coarse-grained transition rates among them~\cite{Battle2016}. By following the same procedure for computing entropy production rate, we can determine the energy dissipation rate at any coarse-grained level. However, it is known that coarse-graining reduces entropy production~\cite{Busiello2019,Busiello2019a}, which makes it a challenging problem to determine the ``true'' dissipation rate at the microscopic level, which is the free energy cost of maintaining the NESS, from the apparent energy dissipation rate determined at a coarse-grained level.

In our recent work~\cite{Yu2021}, we connected dissipation at different scales by developing a coarse-graining procedure inspired by the real space renormalization group (RG) approach by Kadanoff~\cite{Kadanoff1966, Wilson1975} and applying it to various reaction networks in the general state space that can include both physical and chemical state variables.
We found that the energy dissipation rate satisfies an inverse power law with the coarse-graining scale. 
The scaling exponent was found to be the sum of two contributions---a link exponent characterizing the reaction network structure and a term that is determined by the correlation of stationary probability fluxes (currents) in the network. While the link exponent can be calculated directly from the network structure and the coarse-graining process, it proves more difficult to obtain the correlation between the steady-state fluxes, which depends on all the rate constants in the entire network in a nonlinear fashion~\cite{Schnakenberg1976} and also evolves under the coarse-graining process. In our previous work~\cite{Yu2021}, the flux correlation function was computed numerically.

In this work, we derive and solve the functional RG equation of the flux correlation function exactly for hypercubic lattice networks in arbitrary dimensions. 
The hypercubic lattice structure is not only theoretically more tractable, but also represents the state-space structure of a large class of reaction systems, such as chemcial reaction systems including the Brusselator and active matter systems including the MT-kinesin active flow system~\cite{Yu2021} and active disordered media~\cite{Cocconi2021}.
First, we derive the iterative RG equation of the correlation function of steady-state fluxes by following the state space coarse-graining procedure. Then, we solve the functional RG equation exactly and find a family of fixed point solutions. Next, we determine the stability of these solutions.
Finally, we show how the initial condition in the RG equation, i.e., the correlation function in the fine-grained system, determines the fixed point to which the system converges under the  coarse-graining procedure, and thereby the dissipation scaling exponent. 
We solve the problem in detail in the square lattice network before extending the analysis and solutions to 3D (cubic lattice) and higher dimensions.
Possible connections of our problem in particular the existence of a family of fixed point solutions with different exponents to the Kosterlitz-Thouless phase transition will also be discussed.

\section{Square lattice}
We start by establishing the basic theoretical framework of network coarse-graining and correlation renormalization. 
Consider a general nonequilibrium system whose microscopic configuration is represented by discrete states labeled with $i=1,2,\cdots,n$.
The system evolves following continuous-time Markovian dynamics.
Let $P_i(\tau)$ denote the probability of finding the system in state $i$ at time $\tau$ and $k_{i,j}$ the rate of transition from state $i$ to state $j$. 
The probability distribution evolves following the master equation
\begin{align}
	\dv{P_i(\tau)}{\tau} = \sum_j \qty[k_{j,i}P_j(\tau)-k_{i,j}P_i(\tau)],\quad i=1,2,\cdots,n.
\end{align}

Here, we focus on the systems' properties in the steady state where $\dv{P_i}{\tau}=0$ for all the states.
We define $J_{i,j}=k_{i,j}P_i$ as the steady-state probability flux (current) and $A_{i,j} = J_{i,j}-J_{j,i}$ as the net probability flux. For reaction networks out of equilibrium, the breaking of detailed balance relation leads to nonvanishing net probability fluxes ($A_{i,j}\neq 0$) and constant dissipation of free energy in the steady state. The steady-state energy dissipation (entropy production) rate is given by~\cite{Qian2006,Hill1977}
\begin{align}
	\dot{W} = \sum_{i<j} (J_{i,j}-J_{j,i})\ln \frac{J_{i,j}}{J_{j,i}}.
\end{align}

For extended nonequilibrium systems, the state space contains a large number of discrete states.
Solving for the entire steady-state probability distribution requires knowledge of all the microscopic rate constants, which may be implausible experimentally but also unnecessary for analyzing macroscopic observables.
Instead, we can characterize the steady-state properties with a coarse-grained description of the reaction network, where similar microscopic states are combined into one coarse-grained state.
As the network is coarse-grained, various quantities, such as the steady-state probability distribution of the coarsed-grained states and the effective transition rate constants between these coarse-grained states, must be renormalized accordingly.
We refer to the theoretical framework which relates the properties at different coarse-grained scales as the State-Space Renormalization Group (SSRG) inspired by the real space renormalization group approach pioneered by Kadanoff to study critical phenomenon and scaling behavior in the Ising model~\cite{Kadanoff1966,Wilson1975}.
An important and crucial distinction, however, is that the coarse-graining takes place in the general state space, which could include both chemical and physical state variables, rather than the real space whose state variable is the physical location.

In our previous work~\cite{Yu2021}, a coarse-graining procedure was developed and used to analyze the energy dissipation rate for different reaction networks.
We showed that the dissipation rates at different scales obey a scaling law $\dot{W}_b \sim \dot{W}_0 (n_b/n_0)^{-\lambda}$ where $n_b/n_0$ is the number of microstates in each coarse-grained state, which characterizes the scale change. The scaling exponent is given by $\lambda= d_L- \log_r\qty(1+C^*)$. 
Both $d_L$ and $r$ are associated with the number of links (reactions) and vertices (states) to be combined during the coarse-graining procedure. They depend on the network structure and the coarse-graining procedure but not the rate constants.
In contrast, $C^*$ is the asymptotic nearest-neighbor correlation between those steady-state fluxes that are combined during coarse-graining (CG), which depends on the rate constants in the network. In our previous work~\cite{Yu2021}, we computed this correlation at different scales numerically and found that it converges to a fixed value $C^*$ asymptotically as CG progresses to coarser and coarser scales. This observation and the resulting power-law scaling of the dissipation rate suggested the existence of fixed point(s) in the SSRG dynamics. 

In this paper, we develop the SSRG theory to connect flux correlation functions at different scales.
By investigating the SSRG dynamic analytically, we aim to understand the scaling behaviors of the correlation function and thereby the energy dissipation rate. 
In this section, we first describe the SSRG theory in square lattice, which will be generalized to hypercubic cubic lattices in subsequent sections.

\begin{figure}
	\centering
	\includegraphics[width=0.6\textwidth]{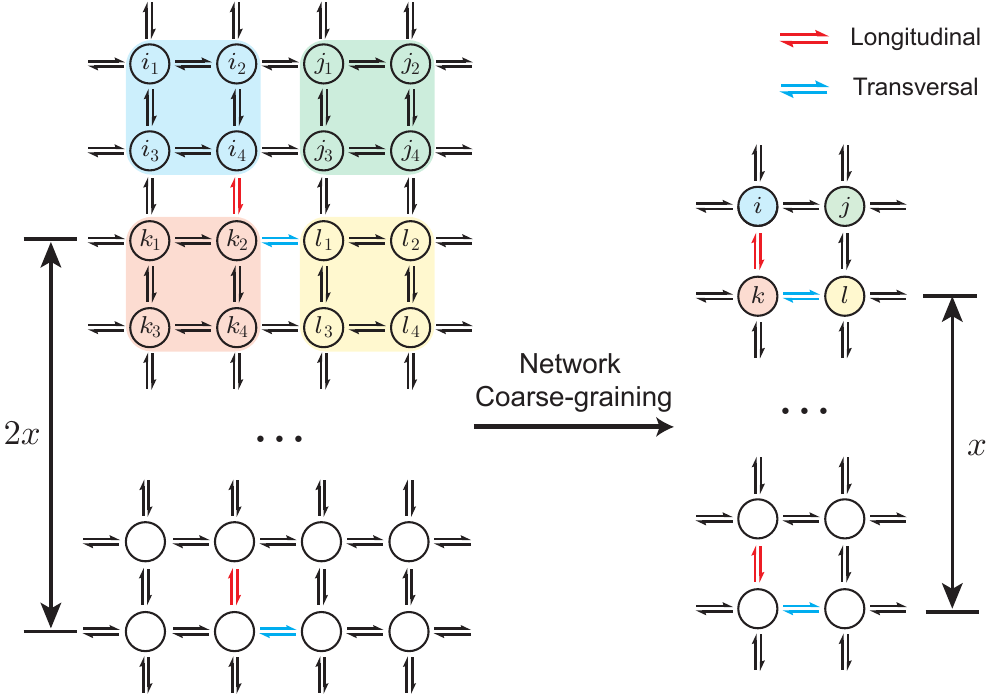}
	\caption{
		An illustration of the coarse-graining of the reaction network and the definition of correlation functions.
		Starting with the network on the left, all the states in the same shaded area are combined together to form a coarse-grained state, which is indicated by the same color on the right. The longitudinal (transversal) correlation function is defined as the correlation between the net fluxes on the red (blue) links, which are parallel (perpendicular) to the direction in which their separation is reduced by half.
	}
	\label{Fig:illustration}
\end{figure}

\subsection{RG equation for flux correlation}
The basic procedure of network coarse-graining of the square lattice is illustrated in Fig.~\ref{Fig:illustration}.
In each iteration, four neighboring states are grouped together to create a coarse-grained state, whose probability is the sum of the probability of all constituent states.
The transition rates in the coarse-grained network are also renormalized to preserve the total (directional) probability flux. For a pair of states $i$ and $j$, for example, the renormalized probability and rate constants are
\begin{equation}
	P_i = \sum_{\alpha=1}^4 P_{i_\alpha},\quad
	k_{i,j} = \frac{J_{i,j}}{P_i}=\frac{k_{i_2,j_1}P_{i_2}+k_{i_4,j_3}P_{i_4}}{P_i}.
\end{equation}
Similar relations exist for other states and other reaction rates.
This process is analogous to the block-spin transformation proposed by Kadanoff~\cite{Kadanoff1966}.

Next, we define $C_t(x)$ as the correlation function between the net fluxes separated by distance $x$ at the coarse-grained level $t$:
\begin{equation}
	C_t(x) = \frac{\expval{A_t(0)A_t(x)}}{\sqrt{\expval{A_t(0)A_t(0)}\expval{A_t(x))A_t(x)}}},
	\label{Eq:Ct definition}
\end{equation}
where $A_t(0)$ and $A_t(x)$ are two fluxes separated by distance $x$ at coarse-grained level $t$.
As shown in Fig.~\ref{Fig:illustration}, the correlation can be defined in both longitudinal and transversal directions.
The exponent for the energy dissipation scaling only depends on the transversal correlation~\cite{Yu2021}, which relates together fluxes which are perpendicular to the spatial separation $x$ (indicated by the blue links in Fig.~\ref{Fig:illustration}).
The longitudinal correlation, which concerns fluxes parallel to the separation $x$, renormalizes trivially and does not contribute to the scaling exponent.
Thus, the correlations we refer to are all transversal unless specifically stated otherwise. 
Let $A(0)$ and $A(x)$ denote two net fluxes separated by $x$ in the transversal direction. The coarse-grained net fluxes are:
\begin{equation}
	A_{t+1}(0) = A_t(0)+A_t(1),\quad
	A_{t+1}(x) = A_t(2x)+A_t(2x+1).
	\label{Eq:A renorm}
\end{equation}
The renormalized (transversal) correlation is obtained by substituting the flux renormalization relations (Eq.~\ref{Eq:A renorm}) into the correlation function definition (Eq.~\ref{Eq:Ct definition}):
\begin{align}
	C_{t+1}(x) = \frac{\expval{A_{t+1}(0)A_{t+1}(x)}}{\sqrt{\expval{A_{t+1}(0)A_{t+1}(0)}\expval{A_{t+1}(x))A_{t+1}(x)}}}
	= \frac{2C_t(2x)+C_t(2x-1)+C_t(2x+1)}{2\qty[1+C_t(1)]}.
	\label{Eq:Def RG Equation}
\end{align}
This is the iterative RG equation for the flux correlation function.
$C_t(x)$ obeys the normalization condition $\sum_{x=-\infty}^{+\infty}C_t(x)=0$ due to the steady-state flux conservation condition. Given that $C_t(-x)=C_t(x)$ and $C_t(0)=1$, the normalization condition can also be written as
\begin{align}
	\sum_{x=1}^{+\infty} C_t(x) = -\frac{1}{2},
	\label{Eq:2d normalization}
\end{align}
which is indeed preserved by the RG iteration equation (Eq.~\ref{Eq:Def RG Equation}). The asymptotic value of the nearest-neighbor flux correlation that determines the scaling exponent for the energy dissipation rate is given by:   $C^*=\lim\limits_{t\to+\infty} C_t(1)$.

The functional RG equation (Eq.~\ref{Eq:Def RG Equation}) is the first key result of this paper. It allows us to predict the flux correlation function at any given scale from the original correlation function at the microscopic scale. Next, we solve Eq.~\ref{Eq:Def RG Equation} to obtain its fixed point solution(s).

\subsection{RG Fixed points}
Despite the general difficulty to identify the fixed points of functional RG equations, which involve infinite degrees of freedom, the fixed points of Eq.~\ref{Eq:Def RG Equation} can be determined exactly. Recognizing that the right hand side is reminiscent of the discrete Laplacian, we introduce two auxiliary functions which are the discrete integral of the correlation function:
\begin{align}
	S_t(x) = \sum_{y=x}^{+\infty}C_t(y),\quad
	T_t(x) = \sum_{y=x+1}^{+\infty} S_t(y).
\end{align}
The normalization condition translates to $S_t(1)=-\frac{1}{2}$, which is decoupled from other degrees of freedom, rendering $S_t(2)$, $S_t(3)$, and so on independent variables.
The RG equations for $S$ and $T$ are:
\begin{align}
	S_{t+1}(x) = \frac{S_t(2x-1)+S_t(2x)}{1-2S_t(2)},\quad
	T_{t+1}(x) = \frac{T_t(2x)}{1-2S_t(2)}.
	\label{Eq:2d ST RG equation}
\end{align}
The $T$ equation implies that at the fixed point (indicated by a superscript $\star$), the ratio $T^\star(x)/T^\star(2x)$ is a constant independent of $x$.
Hence, the equation only admits a power-law solution $T^\star(x)=T^\star(1)x^{-c}$ and a trivial solution $T^\star(x)=0$, which corresponds to two classes of fixed points.
The trivial solution corresponds to a single fixed point, which we label with subscript $0$:
\begin{align}
	C_0^\star(x) = \begin{cases}
		-\frac{1}{2},\quad & x=1     \\
		0,\quad            & x\geq 2
	\end{cases}.
	\label{Eq:2d fixed point 0}
\end{align}
At this fixed point, any long-range correlation is absent beyond the nearest neighbor.

\begin{figure}
	\centering
	\includegraphics[width=0.75\textwidth]{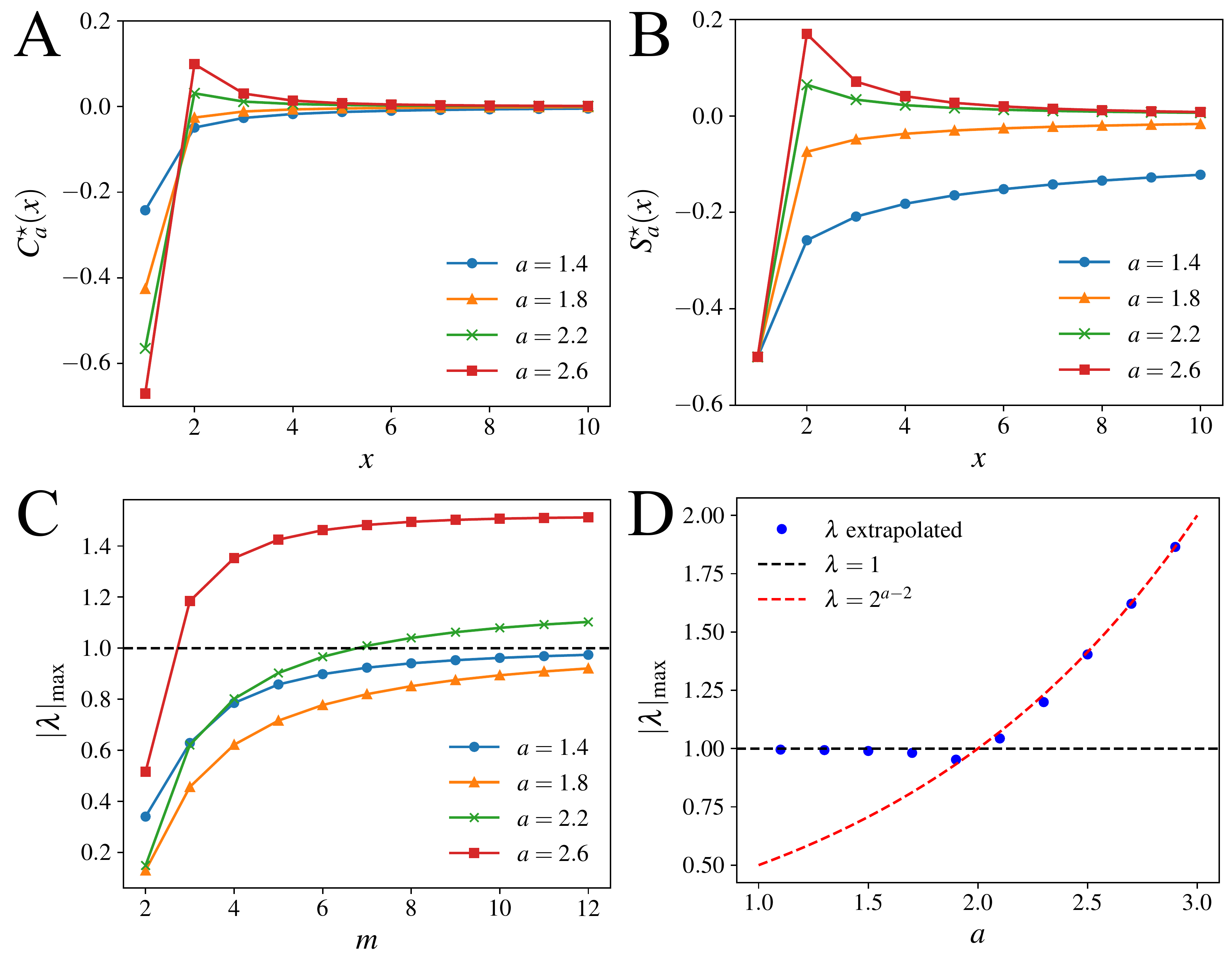}
	\caption{
		Fixed points and linear stability analysis in the square lattice.
		(A--B) Profiles of the correlation $C_a^\star(x)$ and its cumulative sum $S_a^\star(x)$ at the power-law fixed point for different exponents $a$.
		(C) The spectral radius of the Jacobian matrix truncated at dimensions $(2^m-1)$ for different $a$. The black dashed line indicates $\abs{\lambda}_\mathrm{max}=1$, which is the boundary separating stable and unstable fixed points.
		(D) The spectral radius extrapolated to infinite $m$ for different exponents $a$, compared with theory (black dashed line for $a<2$ and red dashed line for $a>2$). 
	}
	\label{Fig:squareLattice}
\end{figure}

The power-law solution $T^\star(x)=T^\star(1)x^{-c}$ corresponds to a family of solutions with different exponents. At long distance (large $x$), $T^\star(x)$ behaves asymptotically as $x^{-c}$; hence $C^\star(x)$, which is the second-order finite difference of $T^\star$, decays as $x^{-c-2}$. We introduce a new exponent $a=2+c$ to better characterize the asymptotic behavior of the correlation itself.
$T^\star(1)$ is obtained by examining the RG equation for $x=1$:
\begin{equation}
	2^{a-2} = \frac{T^\star(1)}{T^\star(2)} = \frac{1}{1-2S^\star(2)}=\frac{1}{1-2T^\star(1)+2T^\star(2)}=\frac{1}{1-2T^\star(1)\qty(1-2^{-a+2})},
\end{equation}
which leads to $T^\star(1) = \frac{1}{2}$ and therefore, $T_a^\star(x) = \frac{1}{2}x^{2-a}$.
Performing finite difference once gives $S_a^\star(x) = \frac{1}{2}[(x-1)^{2-a}-x^{2-a}]$. Another finite difference gives the correlation function:
\begin{equation}
	\begin{aligned}
		C^\star_a(x) = \begin{cases}
			2^{1-a}-1,\quad x=1 \\
			\frac{1}{2}\qty[ (x-1)^{2-a} + (x+1)^{2-a}]-x^{2-a}, \quad x\geq 2
		\end{cases}
	\end{aligned}
	\label{Eq:2d fixed point a}
\end{equation}
where the subscript $a$ labels the asymptotic decay rate for this family of fixed points.  For $a\leq 1$, $C^\star_a$ is indeed a fixed point mathematically, but it does not correspond to the correlation function of any reaction network since the summation of all correlation is divergent, in violation of the normalization condition Eq.~\ref{Eq:2d normalization}. This class of solutions is irrelevant for flux correlation. The $a=2$ solution is also irrelevant since the asymptotic power-law decay breaks down as $c=a-2=0$, resulting in a singular behavior.
Thus, only exponents $a\in(1,2)\cup(2,+\infty)$ correspond to meaningful fixed points of the flux correlation, which can be directly verified by inserting these solutions to the RG equation (Eq.~\ref{Eq:Def RG Equation}).
The profiles of these fixed point solutions are shown in Fig.~\ref{Fig:squareLattice}A\&B.

In summary, the RG equation admits only two types of fixed points: a trivial solution $C_0^\star$ with only nearest neighbor correlation and a family power-law solutions $C_a^\star$ ($a\in(1,2)\cup(2,+\infty)$) with long-range correlation that decays as $x^{-a}$.
These fixed point solutions are presented explicitly in Eq.~\ref{Eq:2d fixed point 0} and Eq.~\ref{Eq:2d fixed point a}, which are the second key result of this paper. 
Next, we study the stability and selection of these fixed point solutions by analyzing the RG dynamics.

\subsection{Stability of the fixed points}
The stability of the fixed points of the correlation function may be analyzed by linear stability analysis, which evaluates the spectral radius of its Jacobian matrix at the fixed point. However, this analysis can not be directly applied to the correlation $C(x)$ itself since they are not independent variables due to the normalization constraint.
Instead, we can study the Jacobian of the cumulative sum $S(x)$, which are independent variables for $x=2,3,\dots$.
The Jacobian of $S$ is defined as follows:
\begin{align}
	{\mathcal{J}} = \pdv{S_{t+1}}{S_t}
	= \pdv{\qty[S_{t+1}(2),S_{t+1}(3),S_{t+1}(4),\cdots]}{\qty[S_{t}(2),S_{t}(3),S_{t}(4),\cdots]}.
\end{align}
For the specific power-law fixed point with exponent $a$, the matrix elements are
\begin{equation}
	{\mathcal{J}_{i,j}} = \eval{\pdv{S_{t+1}(i)}{S_t(j)}}_{S^\star_a} = \frac{\delta_{2i-1,j}+\delta_{2i,j}}{1-2S^\star_a(2)} + \frac{2S_a^\star(i)}{1-2S_a^\star(2)}\delta_{j,2} = 2^{a-2}\qty(\delta_{2i-1,j}+\delta_{2i,j}+2S_a^\star(i)\delta_{j,2}).
\end{equation}
To obtain the spectral radius of this infinite-dimensional matrix, we calculate the eigenvalues with a finite-dimension truncation before taking the dimension to infinity.
Specifically, truncation is made at dimensions $\qty(2^{m}-1)$, which enables the standard Gaussian elimination method to reduce the matrix $\qty(\lambda I_m - J_m)$ to a lower diagonal matrix, revealing the characteristic polynomial:
\begin{align}
	p_m(\lambda)=  \det\qty(\lambda I_m - J_m)
	 & = \lambda^{2^m-(m+1)}\qty(\lambda^m - \qty(\xi^{-1}-1) \sum_{k=0}^{m-1}\lambda^{m-1-k}) = \lambda^{2^m-(m+1)} q_m(\lambda),
\end{align}
where $\xi = 2^{2-a}$, and $I$ is an identity matrix.
The nonzero eigenvalues are simply roots of the polynomial $q_m(\lambda)$. We ask whether $q_m(\lambda)$ has any roots outside the unit disk in the limit of infinite $m$. 
It proves useful to introduce $z=\lambda^{-1}$ and simplify $q_m$ to 
\begin{equation}
	q_{m}(z) = z^{-m}{\qty(1+(1-\xi^{-1})\sum_{k=1}^{m} z^{k})} = {z^{-m}}{\psi_m(z)}.
\end{equation}
To identify all the unstable eigenvalues, we simply look for zeros of $\psi_m(z)$ inside the unit disk. In the infinite $m$ limit, the summation in $\psi_m(z)$ is convergent inside the unit disk, leading to: 
\begin{equation}
	\psi(z) =  \lim\limits_{m\to \infty} \psi_m(z) = \frac{1-z\xi^{-1}}{1-z},\quad \abs{z}<1.
\end{equation}
For $a>2$ and $\xi=2^{2-a}<1$, $\psi(z)$ has a single root $z=\xi$ which is inside the unit disk with all the other roots residing outside the unit disk. In this case, the fixed point is unstable, and there is only one unstable direction with eigenvalue $\lambda = \xi^{-1} = 2^{a-2}>1$.
For $1<a<2$, $\psi(z)$ has no zeros inside the unit disk, and the fixed point is stable. Note that the above analysis could not be directly used to find the eigenvalues associated to the stable directions since the calculation is valid only for $\abs{z}<1$ or equivalently $\abs{\lambda}>1$, but this analysis is sufficient for determining whether the fixed point is stable.

The spectral radius of the Jacobian matrix is evaluated numerically with different truncations (Fig.~\ref{Fig:squareLattice}C). It increases monotonically with $m$ but quickly saturates. The two curves with exponents $a>2$ goes above $1$, clearly indicating that these fixed points are unstable. To extend the results to the infinite system, we extrapolate by fitting the spectral radius $\abs{\lambda}_\mathrm{max}$ evaluated at $m=10,11,12$  to a usual exponential decay function $\lambda(m) = \lambda_\infty-ce^{-\kappa m}$, where $\lambda_\infty$ is the spectral radius of the infinite system. 
The result is shown in Fig.~\ref{Fig:squareLattice}D. The spectral radius stays below $1$ for $a<2$ and reaches $2^{2-a}$ for $a>2$, in complete agreement with the theoretical result. Note that for $a<2$, $\lambda(m)$ approaches $1$ from below but always stays below $1$. Its apparent decrease as $a$ approaches $2$ is due to slower convergence (small $\kappa$). 
These results both numerically analytically support the conclusion that the power-law fixed point is be stable for $1<a<2$ and unstable for $a>2$.

For the trivial fixed point $S^\star(x) = -\frac{1}{2}\delta_{x,1}$, the Jacobian matrix is singular, with the characteristic polynomial given by $p_m(\lambda)=\lambda^{2^m-1}$. We will determine the relevance of this fixed point by directly analyzing the RG flows.

\subsection{Dynamics of the RG flow}
We start by analyzing the asymptotic behavior of the long-range correlation under coarse-graining. Suppose the fine-grained correlation $C_0(x)$ decays asymptotically as $x^{-a_0}$ ($a_0$ is infinity if the decay is faster than power law). The exponent $a_0$ can be obtained from $T_0(x)$:
\begin{equation}
	a_0 = 2 +  \lim\limits_{x\to\infty}\log_2 \frac{T_0(x)}{T_0(2x)}.
\end{equation}
The same exponent can also be evaluated for the renormalized correlation, which we call $a_t$. From the RG equation of $T(x)$, however, we find this exponent to be invariant under the RG iteration:
\begin{equation}
	a_{t+1} = 2 +  \lim\limits_{x\to\infty}\log_2 \frac{T_{t+1}(x)}{T_{t+1}(2x)} = 2 +  \lim\limits_{x\to\infty}\log_2 \frac{T_{t}(2x)}{T_{t}(4x)} = a_t = a_0.
\end{equation}
Therefore, the correlation function converges to either the power-law fixed point with $a=a_0$ or the trivial fixed point (which has no long range correlation and could thus have arbitrary $a_t$).
This observation greatly simplifies the problem as we are now only concerned with the RG flow in the one-dimensional manifold connecting these two fixed points, with all other power-law fixed points rendered irrelevant. The stability analysis in the last section indicates that the RG flows converge to the power-law fixed point when $1<a<2$ and to the trivial fixed point otherwise. To test this, we consider the RG flow starting from points on the straight line connecting these two points, namely:
\begin{equation}
	S_t(x) = p_t S_a^\star(x) + (1-p_t)S_0^\star(x) = p_t S_a^\star(x) = \frac{p_t}{2}\qty[(x-1)^{2-a}-x^{2-a}],\quad x\geq 2.
	\label{Eq:square_lattice_one_solution}
\end{equation}
Substituting this into the RG equation:
\begin{equation}
	S_{t+1}(x) =\frac{S_t(2x-1)+S_t(2x)}{1-2S_t(2)} 
	= p_{t+1}S^\star_a(x) + (1-p_{t+1})S_0^\star(x),
\end{equation}
which leads to the RG equation for $p_t$:
\begin{equation}
	p_{t+1} = \frac{2^{2-a}p_t}{1-(1-2^{2-a})p_t}
	\Rightarrow p_{t+1}^{-1} -1 = 2^{a-2}(p_t^{-1}-1).
	\label{Eq:square_lattice_pt}
\end{equation}
This recursive relation suggests exponential convergence to one of the fixed points. For $a<2$, $p_t$ tends to $1$, indicating convergence to the power-law fixed point. Conversely, $p_t$ tends to $0$ for $a>2$ which represents convergence to the trivial fixed point. For $a=2$, the power-law fixed point does not exist so the correlation always converges to the trivial solution.
Thus, the long-time behavior of the correlation function under RG is completely determined by the asymptotic decay of the initial (fine-grained) correlation function.

\begin{figure}
	\centering
	\includegraphics[width=0.7\textwidth]{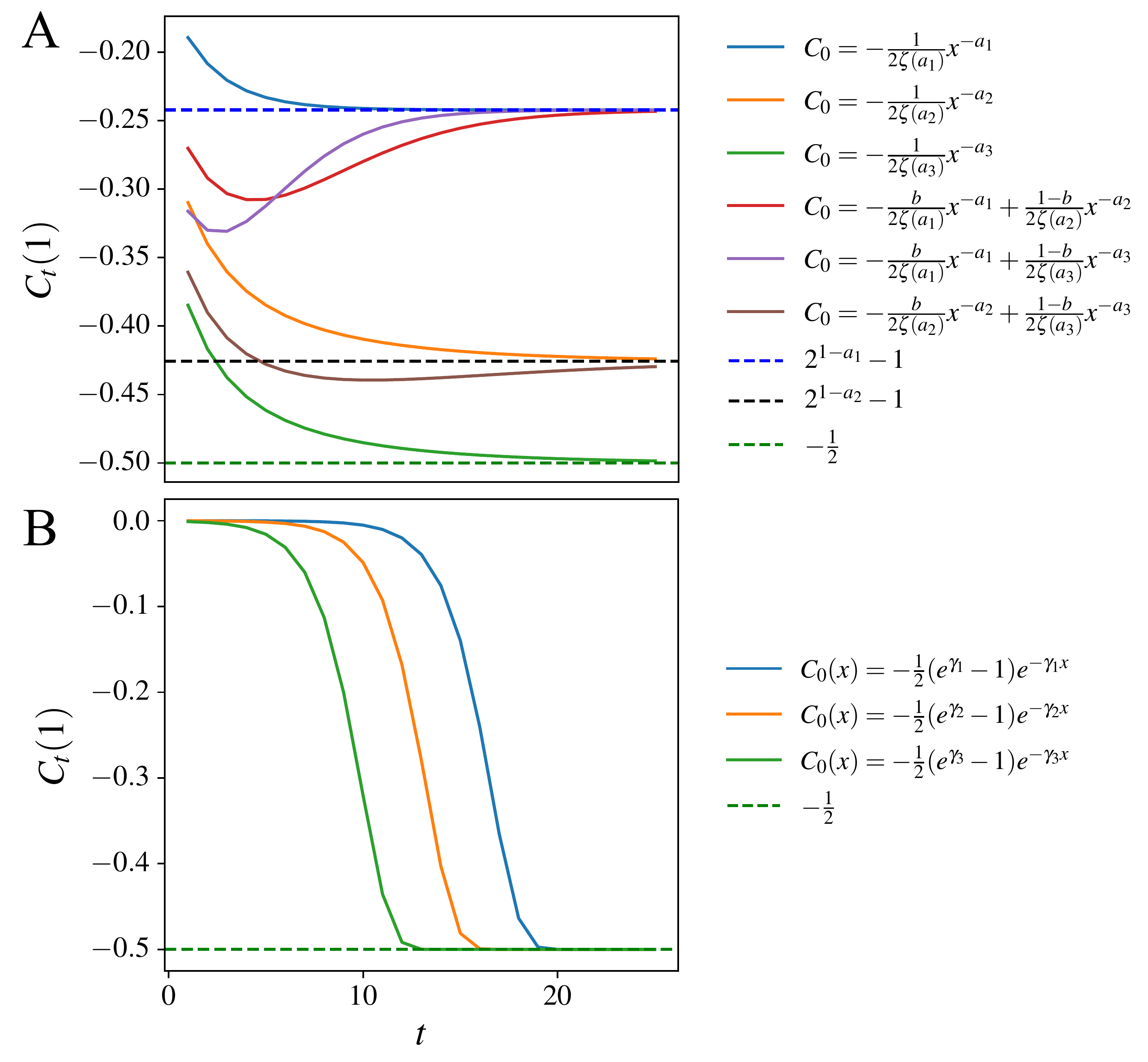}
	\caption{
		The evolution of nearest-neighbor correlation $C_t(1)$ under RG iterations, starting from power-law initialization (A) or exponential initialization (B). 
		The legends of solid lines specify initial conditions $C_0(x)$, and broken lines indicate the values at fixed points.
		For power-law initialization (A), the exponents are $a_1 = 1.4$, $a_2=1.8$, $a_3 = 2.2$. The mixing ratio is $b=0.3$.
		For exponential initialization (B), the decay rates are $\gamma_1 = 10^{-5}$, $\gamma_2 = 10^{-4}$, $\gamma_3 = 10^{-3}$.
	}
	\label{Fig:squareLatticeFlow}
\end{figure}

To test these results further, we calculate the correlation $C_t(x)$ by numerically iterating the RG equation starting from different initial conditions (initial correlation functions) (Fig.~\ref{Fig:squareLatticeFlow}).
If the initial condition consists of a single power-law decay ($C_0(x) = -\frac{1}{2\zeta(a)}x^{-a}$), $C_t(1)$ converges to the power-law fixed point for $a<2$ (blue and orange lines in Fig.~\ref{Fig:squareLatticeFlow}A) and to the trivial fixed point for $a\geq2$ (green line in Fig.~\ref{Fig:squareLatticeFlow}A).
Moreover, the trivial fixed point is always reached if the correlation decays faster than power law. As shown in Fig.~\ref{Fig:squareLatticeFlow}B, $C_t(1)$ always tends to $-\frac{1}{2}$ if the initial correlation is exponential, regardless of how slow the decay is.
These numeric results confirm that $a_c=2$ is the critical point for the stability of the power-law fixed points.

\subsection{Selection of fixed point solutions}
So far, the stability of individual power-law fixed point solutions have been shown with two independent approaches: computing spectral radius of the Jacobian or analyzing the RG flow between power-law and trivial fixed points. 
Due to the existence of a family of power-law solutions, it is natural to draw connection with various pattern/front selection problems~\cite{Cross1993,VanSaarloos1988,VanSaarloos2003} and ask what happens if we start with a combination of multiple solutions. 
In this case, the correlation is dominated by the largest exponent at short distance and the smallest exponent at long distance. After RG, the short-distance variations are washed out and the long-range behavior, which only depends on the smallest exponent, remains and dictates the final correlation profile. Any other terms with faster asymptotic decay is irrelevant in the RG sense. 

To illustrate this point, we consider a simple case with a combination of two nontrivial solutions: $S_t(x) = p_t S^\star_a(x)+q_t S^\star_b(x)$. The RG equations for $p_t$ and $q_t$ are: 
\begin{equation}
	p_{t+1} = \frac{2^{2-a}p_t}{1-\qty(1-2^{2-a})p_t-\qty(1-2^{2-b})q_t}, \quad
	q_{t+1} = \frac{2^{2-b}q_t}{1-\qty(1-2^{2-a})p_t-\qty(1-2^{2-b})q_t}.
	\label{Eq:twoModeRG}
\end{equation}
Between the two solutions, the one with the smaller exponent always prevails over the other one as the ratio of the two coefficients evolves as $\frac{q_t}{p_t} = 2^{-(b-a)t}\frac{q_0}{p_0}$. In fact, the relative decay rate was already hinted in Eq.~\ref{Eq:square_lattice_pt}, where smaller $a$ leads to faster convergence. 
These recursive relations have three fixed points: $(p_1,q_1)=(0,0)$, which is the trivial fixed point; $(p_2,q_2)=(1,0)$ and $(p_3,q_3)=(0,1)$, which are the power-law fixed points with exponents $a$ and $b$, respectively. 
The stable one among them is the trivial fixed point if both exponents are greater than $2$ or the power-law fixed point with the smaller exponent if it is smaller than $2$.

\begin{figure}
	\centering
	\includegraphics[width=0.75\textwidth]{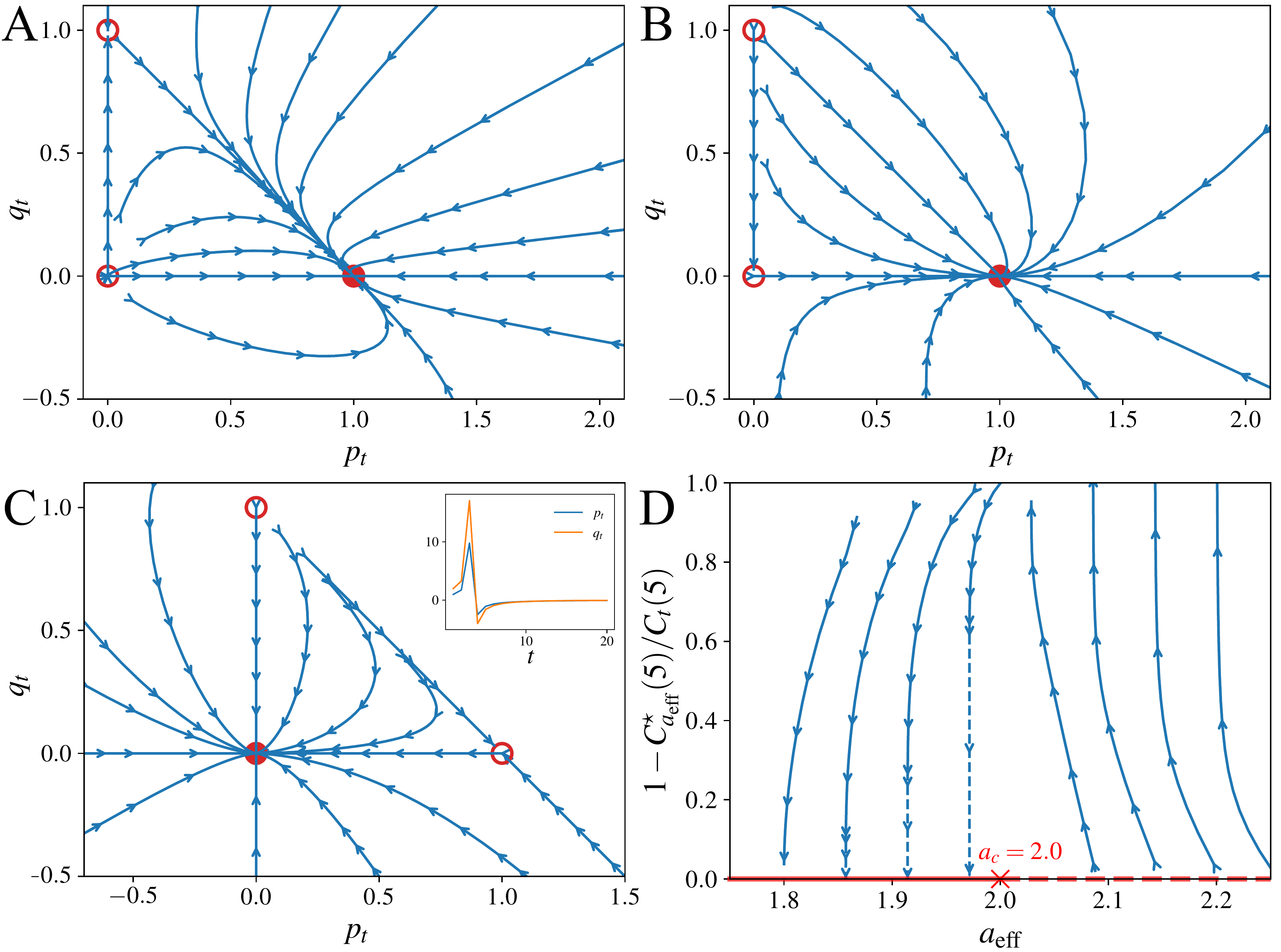}
	\caption{
		RG flows for the square lattice model. 
		(A--C) RG flows in the $(p_t,q_t)$ plane for systems with a combination of two nontrivial solutions (Eq.~\ref{Eq:twoModeRG}) with three different combinations of exponents:
		(A), $a=1.7$, $b=1.8$; (B), $a=1.7$, $b=2.2$; (C), $a=2.2$, $b=2.3$.
		Empty and filled red circles represent unstable and stable fixed points, respectively. 
		Inset of C shows the trajectories starting with $p_t+q_t>1$, which diverge to a large number before converging back to the origin which is the only stable fixed point. 
		(D) RG flows starting with a linear combination of power-law functions. The initial conditions are linear combination of power-laws (i.e., $C_0(x)=b_1x^{-a}+b_2x^{-(a+\Delta a)}$).
		$\Delta a = 0.5$. $b_{1}$ and $b_2$ are chosen to visualize RG flows in a sufficiently large dynamic range. $L=3$ is used to calculate $a_\mathrm{eff}$.
		Solid and broken red lines indicate families of stable and unstable fixed points, and the transition point $a_c=2.0$ is marked with a cross. 
		The solid blue lines show the RG flows from 22 direct RG iterations, and dashed blue lines are extrapolation using the last 6 iterations assuming exponential convergence to a fixed point. 
    }
	\label{Fig:squareLatticeRGFlow}
\end{figure}

In Fig.~\ref{Fig:squareLatticeRGFlow}A, we show the RG flows in the $(p_t,q_t)$ plane for $a<b<2$. While $(1,0)$ is stable and $(0,0)$ is unstable in both directions, $(0,1)$ has both stable and unstable directions. This indicates that the $C_b^\star$ solution is stable by itself but unstable in the presence of another solution with a smaller exponent $a$.
However, the stability changes with exponents.
For $a<2<b$ (Fig.~\ref{Fig:squareLatticeRGFlow}B), the $C_b^\star$ solution becomes unstable in both directions (i.e., even in the absence of $C_a^\star$). $(0,1)$ is completely stable, and $(0,0)$ is stable in the $q_t$ direction but not in the $p_t$ direction.
For $2<a<b$, only the trivial fixed point $(0,0)$ is stable (Fig.~\ref{Fig:squareLatticeRGFlow}C). For the sake of clarity, trajectories with $p_t+q_t>1$ are not shown because they converges to $(0,0)$ after growing to a large value before changing sign (Fig.~\ref{Fig:squareLatticeRGFlow}C, inset). 

Besides direct iteration of Eq.~\ref{Eq:twoModeRG}, we also carry out explicit simulations of the RG dynamics of the correlation function, starting with a linear combination of power-law functions (red, purple, and brown lines in Fig.~\ref{Fig:squareLatticeFlow}A). 
In all the cases examined, the correlation converges to the power-law fixed point corresponding to the smaller exponent, if that exponent is smaller than $a_c=2$.  
Fig.~\ref{Fig:squareLatticeRGFlow}D examines the RG flow, with the $x$ axis quantifying the asymptotic decay with an effective exponent:
\begin{equation}
	a_\mathrm{eff}(t) = \log_2\frac{4C_t(L)}{C_t(2L-1)+2C_t(L)+C_t(2L+1)}.
\end{equation}
The rationale behind this definition is both to recover the log-derivative $a_\mathrm{eff} \sim -\dv{\log{\abs{C}}}{\log{x}}$in the large $L$ limit and to recover the exact value of $a$ in the large $t$ but finite $L$ limit. The $y$ axis is some measure of $(1-p_t^{-1})$ which, as shown in Eq.~\ref{Eq:square_lattice_pt}, vanishes for stable fixed points. The direction of RG flows indicates that there is a family of fixed points along the $x$ axis, which are stable for $a<a_c=2$ and unstable otherwise, in full agreement with theory. The stability and selection of the fixed point solutions summarized as the RG flow diagrams shown in Fig.~\ref{Fig:squareLatticeRGFlow} is the third key result of this paper.

\subsection{Possible connection to Kosterlitz-Thouless transition}

The existence of a continuous line of fixed points and the corresponding RG flow structure in SSRG are reminiscent of the well-known Kosterlitz-Thouless (KT) transition of the two-dimensional XY model~\cite{Berezinskii1972,Kosterlitz1973,Kosterlitz1974}. 
More specifically, analogies can be drawn from the exponent $a$ and the amplitude term $(1-p_t^{-1})$ to the two key parameters characterizing KT transition: the inverse interaction parameter $K^{-1}$ and fugacity $y_0$ of topological defects. 
$y_0$ captures the energy cost of introducing new defects, and $K^{-1}$ characterizes the attraction and repulsion between defects. The recursive relations of $(K^{-1},y_0)$ admits a family of fixed points with $y_0=0$ and different $K^{-1}$. However, only those fixed points below the critical temperature, namely $K^{-1}<K_c^{-1}$, are stable. An analogy begins to emerge in which the exponent $a$ plays a similar role as $K^{-1}$ in characterizing a family of fixed points, and $a_c=2$ is the critical point that corresponds to $K_c^{-1}$, which separates stable fixed points from unstable ones. Moreover, the analogy also extends to how $K^{-1}$ and $a$ both characterizes the exponent of the correlation function in the low-temperature phase (the correlation is given by $\expval{s_0s_r} \propto r^{-K^{-1}/(2\pi)}$ in the low-temperature phase of the two-dimensional XY model, and $\propto r^{-a}$ for the flux correlation). The amplitude term $(1-p_t^{-1})$ plays a similar role as $y_0$, which converges to zero in the low-temperature phase $a<a_c$ and diverges to infinity otherwise.

The striking similarity between the RG flow structure of the two problems suggests that the two systems may be related in a more fundamental way. In the square lattice, the net probability flux is described by a divergence-free flow field, which can be decomposed into cyclic fluxes in the smallest reaction cycles (equivalent to the flow field vorticity in the continuum description). 
These cyclic fluxes (vortices) are analogous to the topological defects in the 2D XY model. However, it is not straightforward to generalize this analogy to hypercubic lattices, whose RG flows, as shown in the following sections, also bear resemblance to the KT transition. It is also important to note that the KT transition takes place in thermodynamic equilibrium while the NESS studied here can be arbitrarily far from equilibrium. Although it may be fruitful to explore further connection between the two problems, we did not manage to find an exact mapping and leave it to future work.

\subsection{Application of the SSRG results}
An important application of the SSRG analysis and the resulting fixed point correlation function is to determine the value of the scaling exponent $\lambda$ for the dissipation rate. For a square lattice where the flux correlation decays as $x^{-a}$, the nearest neighbor correlation at the stable RG fixed point is $C^*=2^{1-a}-1$ for $a<2$ and $C^*=-\frac{1}{2}$ otherwise. The dissipation exponent is
\begin{equation}
	\lambda_{2d} = 1- \log_4(1+C^*) = \begin{cases}
		\frac{1+a}{2},\quad 1<a<2, \\
		1.5,\quad a\geq 2.
	\end{cases}
\end{equation}
In the random flux model where the reaction rates are independent identical and identically distributed random variables (i.i.d.), numeric calculation gives $a=2$, $C^*=-\frac{1}{2}$~\cite{Yu2021}, leading to $\lambda=1.5$ in the infinitely large system. While the analysis here provides theoretical support for previous numeric results in the random flux model~\cite{Yu2021}, it could also be applied to other reaction networks with different reaction rates, which can lead to different exponents ($a$ and $\lambda$).

\section{Cubic lattice}
We now generalize our findings in square lattice to higher dimensions, which constitutes the fourth (and the last) key result of this paper. We start with the cubic lattice where the transversal correlation is denoted by $C_t(x,y)$ as there are now two directions perpendicular to the flux. The correlation is normalized by $\sum_{x,y}C(x,y) = 0$.
The dissipation scaling exponent is
\begin{equation}
	\lambda_{3d} = 1- \log_8\qty(1+C^*),
\end{equation}
where $C^*=\lim\limits_{t\to+\infty} C^*_t=\lim\limits_{t\to+\infty} \qty[C_t(0,1)+C_t(1,0)+C_t(1,1)]$.

\subsection{RG equation and fixed points}
In cubic lattice, each coarse-graining step combines eight adjacent states (vertices) together and merges four parallel reactions (links) into one. The net fluxes are coarse-grained by:
\begin{equation}
	A_{t+1}(x,y) = A_t(2x,2y) + A_t(2x+1, 2y) + A_t(2x, 2y+1) + A_t(2x+1,2y+1),
\end{equation}
where $x,y$ denote the separation in the two transversal directions. The flux correlation $C_t(x,y)$ is renormalized by
\begin{equation}
	\begin{aligned}
		C_{t+1}(x,y) & = \frac{1}{4\left[1+C_t(0,1)+C_t(1,0)+C_t(1,1)\right]} \left[4C_t(2x,2y)\right. \\
		+            & \left. 2C_t(2x-1,2y)+2C_t(2x+1,2y)+2C_t(2x,2y-1)+2C_t(2x,2y+1) \right.          \\
		+            & \left. C_t(2x-1,2y-1)+C_t(2x-1,2y+1)+C_t(2x+1,2y-1)+C_t(2x+1,2y+1) \right]
	\end{aligned}
	\label{Eq:cubicCRecursiveEq}
\end{equation}
This is the iterative RG equation in the cubic lattice. 
Motivated by results in the square lattice, we next look for fixed points by calculating the discrete integrals:
\begin{equation}
	S_t(x,y) = \sum_{u=x}^{+\infty}\sum_{v=y}^{+\infty} C_t(u,v),\quad
	T_t(x,y) = \sum_{u=x+1}^{+\infty}\sum_{v=y+1}^{+\infty} S_t(u,v).
\end{equation}
The recursive relation for $S$ reads:
\begin{equation}
	S_{t+1}(x,y) = \frac{S_t(2x-1,2y-1)+S_t(2x-1,2y)+S_t(2x,2y-1)+S_t(2x,2y)}{4\qty[1+C_t(0,1)+C_t(1,0)+C_t(1,1)]}.
	\label{Eq:cubicSRecursiveRelation}
\end{equation}
Assuming axial reflection symmetry (relaxed below), normalization reads $S_t(0,1)+S_t(1,0)=-\frac{1}{2}$.
The recursive relation for $T$ becomes:
\begin{equation}
	T_{t+1}(x,y) = \frac{T_t(2x,2y)}{4\qty[1+C_t(0,1)+C_t(1,0)+C_t(1,1)]},\quad x^2+y^2>0,
	\label{Eq:cubicTRecursiveRelation}
\end{equation}
which is greatly simplified compared to those for $C$ and $S$.
At the fixed point, the ratio $T^\star(x,y)/T^\star(2x,2y)$ is a constant independent of $x$ and $y$. To see a direct analogy with the square lattice case, we change into polar coordinates with $x=r\cos\theta$ and $y=r\sin\theta$.
Here, the ratio $T^\star(r,\theta)/T^\star(2r,\theta)$ is independent of $r$. Therefore, the radial and angular directions are decoupled, and the radial equation is identical to the $x$ equation in the square lattice. $T^\star$ has nontrivial power-law solutions of the form
\begin{equation}
	T^\star(r,\theta) = f(\theta) r^{4-a},
\end{equation}
where the exponent is denoted as $(4-a)$ so that $a$ directly characterizes the asymptotic decay of $C^\star$ itself. The angular distribution function $f(\theta)$ is a continuous function that captures any anisotropy of the flux correlation.  
The effect of anisotropy will be discussed in detail later as it was not previously studied in the square lattice which has only one transversal direction. For now, it suffice to treat $f(\theta)$ as some arbitrary function with proper normalization. 
At the fixed point, the denominator in Eq.~\ref{Eq:cubicTRecursiveRelation} is given by:
\begin{equation}
	1+C^\star(0,1)+C^\star(1,0)+C^\star(1,1) = \frac{T^\star(2x,2y)}{4T^\star(x,y)} = 2^{2-a}.
	\label{Eq:cubicDenominator}
\end{equation}
The fixed points of $S^\star$ and $C^\star$ are obtained by performing finite difference:
\begin{align}
	S^\star(x,y) = T^\star(x-1,y-1)+T^\star(x,y)-T^\star(x,y-1)-T^\star(x-1,y), \label{Eq:cubicSsolution}
	\\
	C^\star(x,y) = S^\star(x,y) + S^\star(x+1,y+1) - S^\star(x,y+1) - S^\star(x+1,y).
\end{align}
Although their analytic expressions are too complicated to write explicitly (especially due to arbitrariness in the angular distribution), it is clear that asymptotically $S^\star$ decays as $r^{2-a}$ and $C^\star$ as $r^{-a}$. The normalization condition for $C$ requires $a>2$. The above derivation does not cover the correlation near the axes, namely $S(x,0)$, $S(0,y)$, and $S(1,1)$, as $T^\star(x-1,y-1)$ is not defined at these points. These terms are obtained by performing the same analysis separately along the two axes. Along the $y$ axis, for example, $S^t(0,y)$ renormalizes as
\begin{equation}
	S_{t+1}(0,y) = \frac{2\qty[S_t(0,2y-1)+S_t(0,2y)]}{4\qty[1+C_t(0,1)+C_t(1,0)+C_t(1,1)]}, \quad y\geq 1,
\end{equation}
which is different from the recursive relation off-axes (Eq.~\ref{Eq:cubicSRecursiveRelation}) but very similar to that in square lattice. The fixed point solution to this equation is already known from the analysis in 2D:
\begin{equation}
	S^\star(0,y) = \begin{cases}
		-f\qty(\frac{\pi}{2}), \quad y=1 \\
		f\qty(\frac{\pi}{2}) \qty[(y-1)^{3-a}-y^{3-a}],\quad y \geq 2.
	\end{cases}
	\label{Eq:cubicSsolutionY}
\end{equation}
$S^\star$ decays asymptotically as $y^{2-a}=r^{2-a}$ along the $y$ axis, which is consistent with the $r^{2-a}$ decay found off-axes. 
Similar results exist along along the $x$ axis: 
\begin{equation}
	S^\star(x,0) = \begin{cases}
		-f\qty(0), \quad x=1 \\
		f\qty(0) \qty[(x-1)^{3-a}-x^{3-a}],\quad x \geq 2.
	\end{cases}
	\label{Eq:cubicSsolutionX}
\end{equation}
The normalization condition $S_t(0,1)+S_t(1,0)=-\frac{1}{2}$ turns into a constraint on the angular distribution $f(\theta)$:
\begin{equation}
	f(0) + f\qty(\frac{\pi}{2})=\frac{1}{2}.
\end{equation}
Finally, $S^\star(1,1)$ is determined by substituting $C^\star$ with $S^\star$ in Eq.~\ref{Eq:cubicDenominator}:
\begin{equation}
	S^\star(1,1) = \frac{1}{2}-2^{2-a}+S^\star(2,2)-S^\star(0,2)-S^\star(2,0),
\end{equation}
where $S^\star(2,2)$, $S^\star(0,2)$, $S^\star(2,0)$ are given by Eq.~\ref{Eq:cubicSsolution}, Eq.~\ref{Eq:cubicSsolutionY}, and Eq.~\ref{Eq:cubicSsolutionX}, respectively. The set of equations presented here gives the full solution to $S^\star$ as function of $a$ and $f(\theta)$. The flux correlation $C^\star$ can be calculated by finite difference of $S^\star$, with the normalization condition already fulfilled by $f(\theta)$.

In addition to the power-law fixed points described above, there is also a trivial fixed point where $T^\star=0$ except for nearest neighbors. The correlation vanishes except between nearest neighbors, namely $C^\star(0,1)$ and $C^\star(1,0)$. They satisfy normalization condition $C^\star(0,1)+C^\star(1,0) = -\frac{1}{2}$, but the correlation along the two axes need not be equal.

Despite the large degrees of freedom generated by the possibility of having angular anisotropy $f(\theta)$, the dissipation scaling exponent depends solely on the radial asymptotic behavior. 
At the power-law fixed point, Eq.~\ref{Eq:cubicDenominator} leads to $C^* = 2^{2-a}-1$ and $\lambda_{3d,\ \mathrm{power-law}} = \frac{1+a}{3}$.
At the trivial fixed point, the dissipation exponent is $\lambda_{3d,\ \mathrm{trivial}}=\frac{4}{3}$. 

The above analysis indicates that there is a family of power-law fixed points and also a family of trivial fixed points for cubic lattice. This is completely analogous to the fixed point structure for square lattice. The only difference is that each power-law fixed point is now multiplexed by the possibility of having arbitrary angular distributions and that the trivial fixed point is multiplexed by possible asymmetry between the two axis. However, this angular distribution affects neither the scaling exponent or the stability of fixed points, which we show next.

\subsection{Stability}
In the square lattice case, stability of a given fixed point correlation function was analyzed by calculating the spectral radius of the Jacobian of $S^\star$. In fact, $S^\star$ is the only possible candidate for this analysis as $C^\star$ is not an independent variable due to the normalization constraint and $T^\star$ has no closed-form recursive relation. $S^\star(x)$ decays as $x^{1-a}$ in 2D, and the stability condition is $a<2$, or equivalently that $S^\star(x)$ decays slower than $x^{-1}$.
In the cubic lattice, $S^\star(r)$ decays as $r^{2-a}$.
As the angular dependence is decoupled from the radial behavior, the stability problem is effectively one-dimensional, so results similar to those in 2D should be expected in 3D. 
By analogy, it is reasonable to hypothesize that $S^\star(r)$ is also stable only when it decays slower than $r^{-1}$, namely, $2<a<3$ (the lower bound is due to normalization). The hypothesis predicts the critical value $a_c=3$. It is further supported by the continuous transition of the scaling exponent at the critical point. Below $a_c$, the power-law fixed point is stable with exponent $\lambda_{3d,\ \mathrm{power-law}} =\frac{a+1}{3}$.
As $a$ approaches $a_c$, the exponent tends to $\frac{4}{3}$, which is exactly the exponent of the trivial fixed point ($\lambda_{3d,\ \mathrm{trivial}}=\frac{4}{3}$). In other words, there is no sudden jump in the scaling exponent at the transition.
Similar absence of discontinuity in $\lambda$ is also observed in 2D, where $a_c=2$ and $\lambda_c = \frac{3}{2}$.
These evidences suggest that the analogy between square and cubic lattices should extend from the fixed point distributions to their stability.

To make the above analogy more concrete, we study the RG flows in 3D.  
The recursive relation Eq.~\ref{Eq:cubicTRecursiveRelation} suggests that the ratio $T_{t+1}(r,\theta)/T_t(2r,\theta)$ is independent of $\theta$. Hence, $f(\theta)$ is invariant under RG transformations. Similar to the calculation in square lattice, we could define a time-dependent exponent $a_t$ to characterize the asymptotic decay in the radial direction:
\begin{equation}
	a_t = 4 + \lim\limits_{r\to\infty} \log_2 \frac{T_t(r,\theta)}{T_t (2r,\theta)}.
\end{equation}
The exponent is also invariant under RG transformations. Therefore, starting from the fine-grained correlation that decays asymptotically as $r^{-a}$, the system could only converges to either the power-law fixed point with the same exponent $a$ or the trivial fixed point, with the angular distribution $f(\theta)$ unchanged. This observation reduces the problem to a one-dimensional manifold involving only two fixed points.

Following the procedure in 1D, we study the RG flow along the straight line connecting the two fixed points, namely $S_t=p_tS^\star_a+(1-p_t)S^\star_0$. The recursive relation for $S$ reads:
\begin{equation}
	S_{t+1}(x,y) = \frac{S_t(2x-1,2y-1)+S_t(2x-1,2y)+S_t(2x,2y-1)+S_t(2x,2y)}{2-4\qty[S_t(1,1)+S_t(0,2)+S_t(2,0)-S_t(2,2)]},
\end{equation}
Plugging in the expression of $S^\star_a$ yields
\begin{equation}
	p_{t+1} S^\star_a(x,y) = \frac{2^{4-a}p_tS_a^\star(x,y)}{2-4p_t\qty(\frac{1}{2}-2^{2-a})}.
\end{equation}
The RG equation for $p_t$ is
\begin{equation}
	p_{t+1}^{-1} -1 = 2^{a-3}\qty(p_t^{-1}-1),
\end{equation}
which reveals the critical exponent $a_c=3$. For $a<a_c$, $p_t$ tends to $1$ and the power-law fixed point is stable and selected. For $a>a_c$, $p_t$ vanishes, indicating that the power-law fixed point is unstable, and the trivial fixed point is selected instead.
Notably, the RG equation for $p_t$ is identical to that in 2D except for a different $a_c$. The scaling exponent in 3D is
\begin{equation}
	\lambda_{3d} = 1- \log_8(1+C^*) = \begin{cases}
		\frac{1+a}{3},\quad 2<a<3, \\
		\frac{4}{3},\quad a\geq 3.
	\end{cases}
\end{equation}
which depends only on a single parameter $a$ that characterizes the asymptotic behavior of the fine-grained correlation.

\subsection{Numerical test with isotropic correlations}
\begin{figure}
	\centering
	\includegraphics[width=0.7\textwidth]{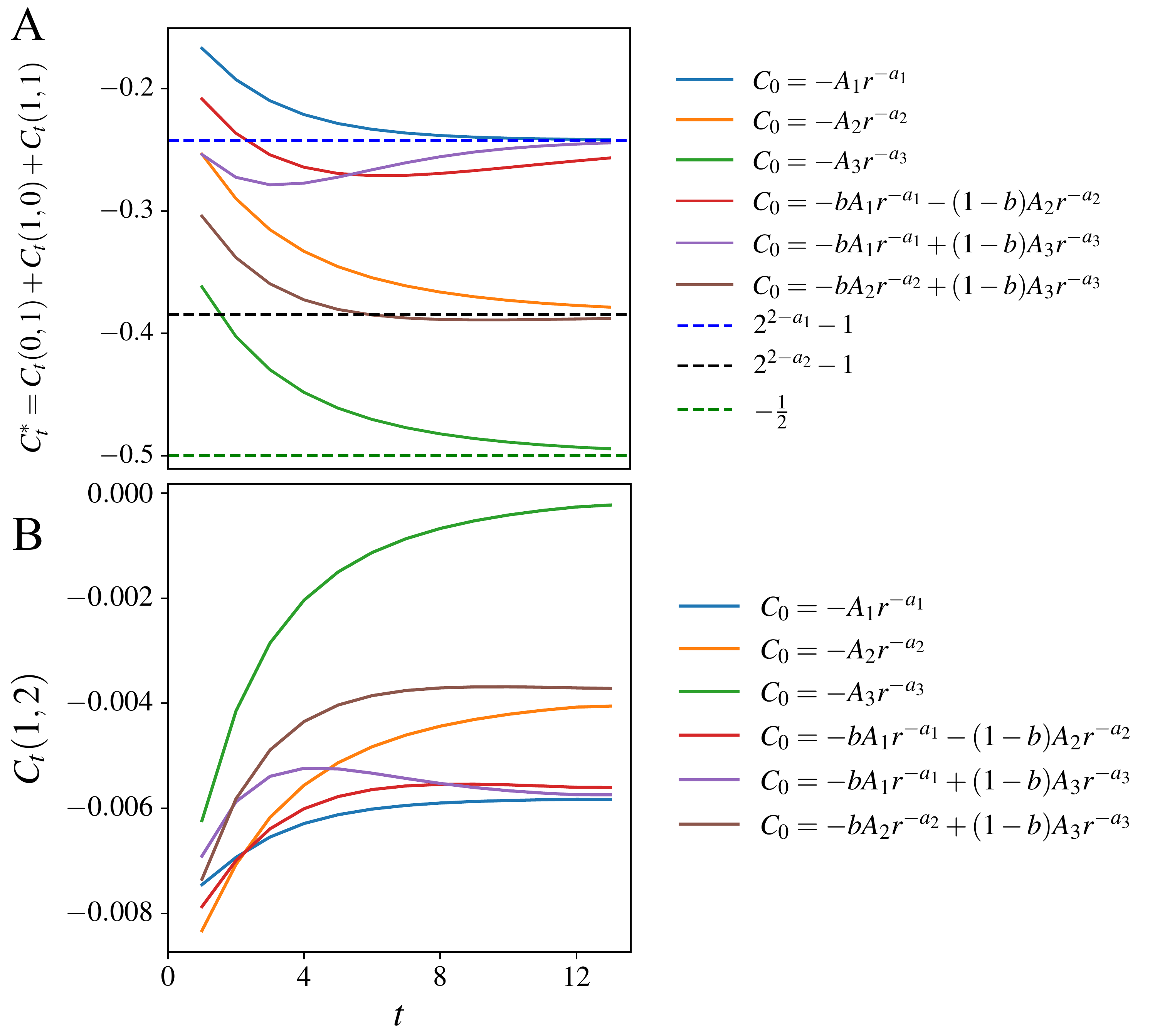}
	\caption{
    	RG iterations in the cubic lattice network with isotropic initialization.
    	(A) Dynamics of $C^*_t=C_t(0,1)+C_t(1,0)+C_t(1,1)$, which determines the dissipation scaling exponent $\lambda_{3d}$.
    	(B) Dynamics of a non-neighboring correlation $C_t(1,2)$, which captures the behavior of correlation at generic separations.
    	The legends of solid lines specify initial conditions with $A_{1,2,3}$ being the proper normalization constants. 
    	The dashed lines indicate values at the fixed point.
    	The exponents are $a_1 = 2.4$, $a_2=2.7$, $a_3 = 3.3$.	
    	The mixing ratio is $b=0.5$.
    	Note that the curves that come together in panel A also converge in panel B.
	}
	\label{Fig:cubicLatticeIsotropic}
\end{figure}

To test the analytic results, we numerically conduct the RG iterations of the correlation with different initial conditions (fine-grained correlations).
We start with the isotropic case where the correlation only depends on the separation distance $r$. The simplest case is a single power law, namely $C_0(x,y) = -A_a r^{-a}$ where $A_a$ ensures proper normalization. Fig.~\ref{Fig:cubicLatticeIsotropic}A demonstrates that the $C^*$ (and therefore the dissipation exponent) converges to the power-law fixed point for $a<3$ (blue and orange curves) and to the trivial fixed point for $a>3$ (green curve). We also examine initialization with a linear combination of power laws. The same figure demonstrates that the smaller exponent always dictates the long-time behavior (red, purple, and brown curves), confirming the physical picture that only the asymptotic behavior determines the long-time behavior, which is unaffected by the addition of any faster-decaying terms.

Notably, rather than being limited to $C^*$ only, the convergence discussed here extends to the entire correlation function $C(x,y)$. For example, Fig.~\ref{Fig:cubicLatticeIsotropic}B plots a non-neighboring correlation $C_t(1,2)$ for the cases studied in panel A. The blue, red, and purple curves, which share the same dominating exponent $a_1$, come together after very different trajectories. The same behavior is observed for other exponents (e.g. see orange and brown curves with the same exponent  $a_2$), and also for correlation at generic separations.  
The parallel between $C^*$ convergence and $C_t(x,y)$ convergence discussed here depend on all the curves having the same $f(\theta)$, which is not the case in the anisotropic cases studied next. 

\subsection{The effect of anisotropy}
\begin{figure}
	\centering
	\includegraphics[width=0.75\textwidth]{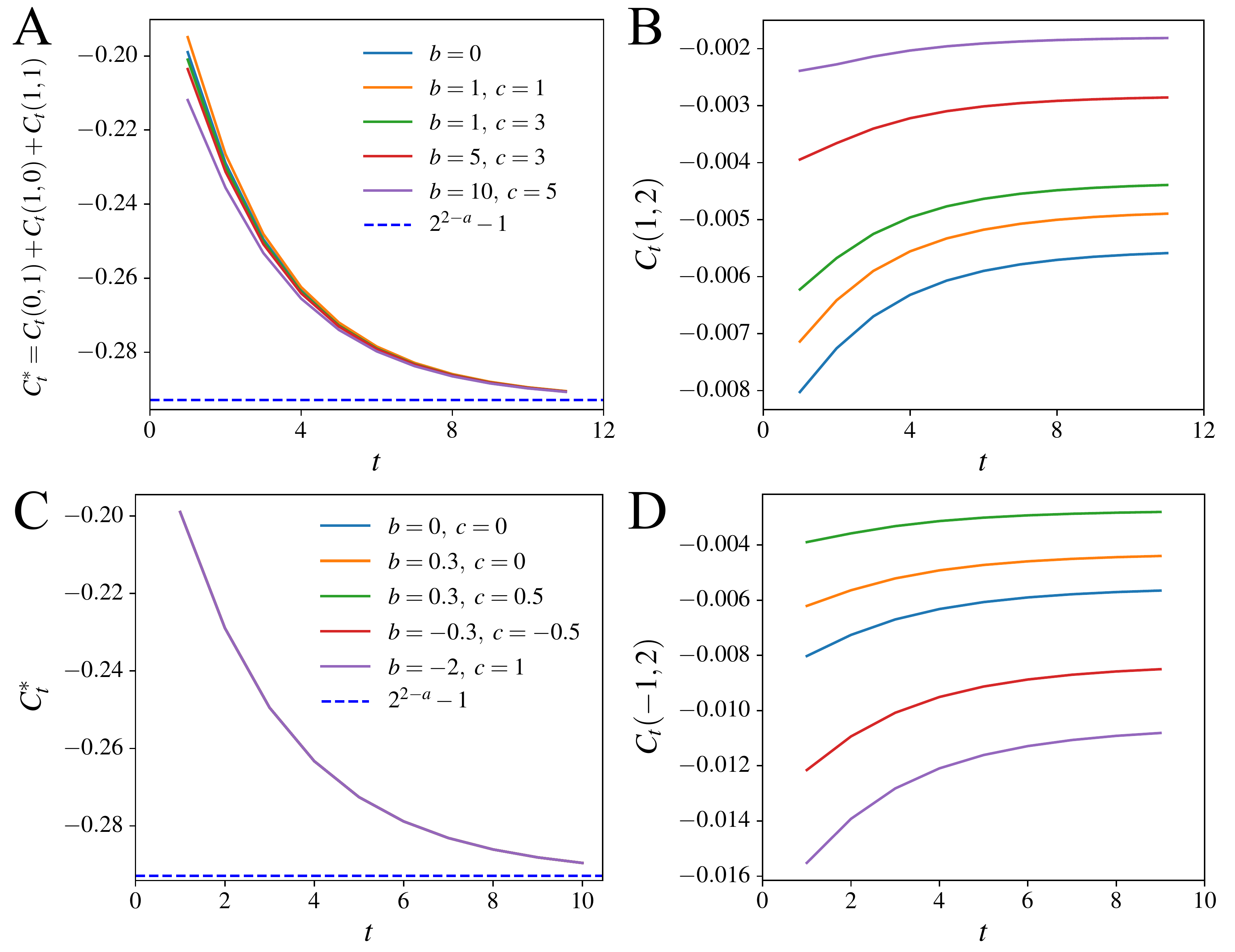}
	\caption{
	    RG iterations in the cubic lattice network with initial conditions that break different symmetries.
	    (A--B) start with $f(\theta) = f_0(1+b\abs{\cos\theta}^c)$, which breaks isotropy.
	    (C--D) start with $f(\theta) = f_0 \qty[1+ b\sin(2\theta) + c \cos(2\theta)]$, which breaks axial reflection symmetry. 
    	(A) Dynamics of $C^*_t=C_t(0,1)+C_t(1,0)+C_t(1,1)$, which determines the dissipation scaling exponent $\lambda_{3d}$.
    	(B) Dynamics of a non-neighboring correlation $C_t(1,2)$, which captures the behavior of correlation at generic separations.
    	(C) Dynamics of the generalized $C^*_t$ (defined in Eq.~\ref{Eq:cubicCRecursiveDenominator}).
    	(D) Dynamics of a non-neighboring correlation $C_t(-1,2)$.
		In all cases,  the exponent is $a = 2.5$, and the values of $b$ and $c$ are given in the legends. $f_0$ ensures proper normalization. 
		Panels (A) and (B) share the same legends, and so do panels (C) and (D). 
		The dashed lines indicate values at the fixed point.
		Note that the curves converge in panels (A) and (C) but not in panels (B) or (D).
    }
	\label{Fig:cubicLatticeAnisotropicAxial}
\end{figure}

A major difference between square and cubic lattices is anisotropy $f(\theta)$ which is only possible with multiple transversal directions present. 
To probe its impact on the RG behavior, we consider initialization with $C_0(r,\theta) = - f(\theta) r^{-a}$, with angular distribution $f(\theta ) = f_0(1+b\abs{\cos\theta}^c)$.
$f_0$ ensures proper normalization, and $b,c$ characterize anisotropy intensity. 
Fig.~\ref{Fig:cubicLatticeAnisotropicAxial}A\&B present the numeric results for $a=2.5$ and different choices of $b$ and $c$.
As shown by Fig.~\ref{Fig:cubicLatticeAnisotropicAxial}A, $C^*_t$ converges to the same value ($C^*_a=2^{2-a}-1$, blue dashed line), regardless of the different anisotropic terms in the initial conditions. 
However, the entire correlation profile definitely does not converge, as shown by different values of $C_t(1,2)$ in Fig.~\ref{Fig:cubicLatticeAnisotropicAxial}B. The lack of profile convergence captures the preservation of $f(\theta)$ during renormalization, but it is not relevant to the calculation of the dissipation exponent $\lambda$. 
To determine $\lambda$, one simply finds the exponent $a$ from the asymptotic behavior and compare it with $a_c$.
These general conclusions do not depend on parameter values or the functional form of $f(\theta$).

\subsection{Axial reflection symmetry}
Having demonstrated the irrelevance of anisotropy in determining the exponent $\lambda$, we now relax the condition of axial reflection symmetry with respect to both axes, which was assumed in the derivation. In the absence of this symmetry, the recursive relation (Eq.~\ref{Eq:cubicCRecursiveEq}) must be modified by replacing the denominator with:
\begin{equation}
	4\qty(1+C_t^*)=
	4+2C_t(1,0)+2C_t(-1,0)+2C_t(0,1)+2C_t(0,-1)+C_t(1,1) + C_t(1,-1)+C_t(-1,1)+C_t(-1,-1),
	\label{Eq:cubicCRecursiveDenominator}
\end{equation}
which gives the generalized definition of $C_t^*$. The full correlation could be decomposed into one symmetric and three anti-symmetric components:
\begin{equation}
	C_t(x,y) = C_t^\mathrm{s}(x,y)+C_t^\mathrm{a1}(x,y)+C_t^\mathrm{a2}(x,y)+C_t^\mathrm{a3}(x,y),
\end{equation}
where
\begin{equation}
	\begin{aligned}
		C_t^\mathrm{s}(x,y)=C_t^\mathrm{s}(-x,y)=C_t^\mathrm{s}(x,-y),     & \quad C_t^\mathrm{a1}(x,y)=-C_t^\mathrm{a1}(-x,y)=C_t^\mathrm{a1}(x,-y),  \\
		C_t^\mathrm{a2}(x,y)=C_t^\mathrm{a2}(-x,y)=-C_t^\mathrm{a2}(x,-y), & \quad C_t^\mathrm{a3}(x,y)=-C_t^\mathrm{a3}(-x,y)=-C_t^\mathrm{a3}(x,-y).
	\end{aligned}
\end{equation}
The decomposition exists and is unique for any correlation function $C_t(x,y)$. In the recursive relation for $C$ (Eq.~\ref{Eq:cubicCRecursiveEq}), all the anti-symmetric components cancel out in the denominator (Eq.~\ref{Eq:cubicDenominator}), and the numerator is a linear combination of both symmetric and anti-symmetric components.
The recursive relations could be established for these components separately by applying the same decomposition to $t$ and $t+1$.
The recursive relation of the anti-symmetric components is completely linear with a normalizing coefficient depending on the symmetric components, while the symmetric components renormalizes following exactly Eq.~\ref{Eq:cubicCRecursiveEq} which was studied in previous sections.

The linearity of the anti-symmetric RG equations ensures that these components do not affect each other or influence the symmetric component, but the symmetric component affects the recursive relation of all components through the denominator.
In other words, the symmetric component affects the renormalization of the anti-symmetric components, but not the other way around.
Since $C^*$ and therefore $\lambda$ only depend on the symmetric components, we can simply ignore the anti-symmetric components and symmetrize the correlation before carrying out any RG analysis:
\begin{equation}
	C_0^\mathrm{s} = \frac{1}{4}\qty[C_0(x,y)+C_0(-x,y)+C_0(x,-y)+C_0(-x,-y)].
\end{equation}

To test the above analysis, we carry out the RG iteration numerically with the following angular distribution:
\begin{equation}
	f(\theta) = f_0 \qty[1+ b\sin(2\theta) + c \cos(2\theta)],
\end{equation}
which no longer has reflection symmetries.
In Fig.~\ref{Fig:cubicLatticeAnisotropicAxial}C\&D, we show results with $a=2.5$ and generic values of $b$ and $c$.  
The systems with different $b$ and $c$ have very different correlation profiles, as exemplified by one of the elements $C_t(-1,2)$ (Fig.~\ref{Fig:cubicLatticeAnisotropicAxial}D). However, $C^*_t$ converges to the same fixed point value $(2^{2-a}-1)$ for these systems (Fig.~\ref{Fig:cubicLatticeAnisotropicAxial}C). This demonstrates that the anti-symmetric components affect the correlation profile but not the exponent $\lambda$.

\section{Hypercubic lattices}
We now generalize the results to hypercubic lattices of arbitrary dimension $n$.  
In 3D, the correlation is first symmetrized to eliminate all the anti-symmetric components. The recursive relation is then simplified by calculating discrete integrals, which decouples radial and angular dependencies to reveal the RG fixed points.
Both symmetrization and discrete integral could be generalized straightforwardly to higher dimensions, which would reduce the $n$-dimensional problem to a one-dimensional problem with the radial asymptotic behavior determining $C^*$ and $\lambda$.
The radial equation admits only power-law solutions $C\sim r^{-a}$, and the critical value of the power exponent is exactly the dimension of the lattice, namely $a_c=n$. The power-law fixed point is stable for $a<a_c$ and unstable for $a>a_c$. As a result, the scaling exponent is given by $\lambda = \frac{1+\min(a,n)}{n}$.

Now we present the mathematical formalism to make these arguments concrete. In $n$-dimensional hypercubic lattice, there are $(n-1)$ transversal directions, so correlation is denoted by $C_t(\mbr)$, where $\mbr = (r_1,r_2,\dots,r_{n-1})$. The symmetrization procedure is given by
\begin{equation}
	C^\mathrm{s}_0(\mbr) = \frac{1}{2^{n-1}} \sum_{s_1,s_2,\dots,s_{n-1}=\pm1} C_0(s_1r_1,s_2r_2,\dots,s_{n-1}r_{n-1}).
\end{equation}
For the sake of simplicity, the superscript $\mathrm{s}$ is dropped, and the correlations are symmetrized unless otherwise stated. The recursive relation reads
\begin{equation}
	C_{t+1}(\mbr) = \frac{1}{2^{n-1}\qty(1+C^*_t)} \sum_{\Delta \mbr_1, \Delta \mbr_2 \in\{0,1\}^{n-1}} C_t(2\mbr+\Delta \mbr_1-\Delta \mbr_2 ),
\end{equation}
where
\begin{equation}
	1+C^*_t= \frac{1}{2^{n-1}}\sum_{\Delta \mbr_1, \Delta \mbr_2 \in\{0,1\}^{n-1}} C_t(\Delta \mbr_1-\Delta \mbr_2 ).
\end{equation}
To simplify the recursive relations, we introduce the discrete integrals
\begin{equation}
	S_t(\mbr) = \sum_{\substack{r_i'=r_i\\ 1\leq i\leq n-1}}^{+\infty} C_t(\mbr'),\quad
	T_t(\mbr) = \sum_{\substack{r_i'=r_i+1\\ 1\leq i\leq n-1}}^{+\infty} S_t(\mbr'),
\end{equation}
whose recursive relations are
\begin{equation}
	S_{t+1}(\mbr) =\frac{1}{2^{n-1}\qty(1+C^*_t)}\sum_{\Delta \mbr_2 \in\{0,1\}^{n-1}} S_t(2\mbr-\Delta \mbr_2 ),\quad
	T_{t+1}(\mbr) =\frac{T_t(2\mbr)}{2^{n-1}\qty(1+C^*_t)}.
	\label{Eq:Hypercubic ST Equations}
\end{equation}
The $T_t(\mbr)$ equation separates radial and angular dependencies (There are now $(n-2)$ angular directions). The equation has both power-law and trivial fixed points. At the power-law fixed point, the angular distribution is arbitrary, and the radial equation has solutions of the form $T_a^\star(r)\propto r^{2(n-1)-a}$ or equivalently $C_a^\star(r)\propto r^{-a}$. The normalization condition requires $a>n-1$. At the trivial fixed point, the correlation is zero except between nearest neighbors ($\abs{\mbr}=1$). 
The dissipation scaling exponent is 
\begin{equation}
	\lambda_{nd} = 1-\frac{1}{n}\log_2(1+C^*).
\end{equation}
At the power law fixed point:
\begin{equation}
	\lambda_{nd,\ \mathrm{power-law}}= 1-\frac{1}{n}\log_2\frac{T_a^\star(2r)}{2^{n-1}T_a^\star(r)} = \frac{a+1}{n}.
\end{equation}
At the trivial fixed point:
\begin{equation}
	\lambda_{nd,\ \mathrm{trivial}}= 1-\frac{1}{n}\log_2\qty(1+C^*) = \frac{n+1}{n}.
\end{equation}
The two exponents meet at $a=n$, which is actually the critical value separating stable and unstable power-law fixed points. To demonstrate this, we carry out stability analysis following methods used in 2D and 3D. Since the asymptotic behavior of $T(r)$ at large $r$ is invariant under RG, a system whose fine-grained correlation decays as $C_0(\mbr) \propto r^{-a}$ could only converge to a power-law fixed point with exponent $a$ or a trivial fixed point. We then study the RG flow direction at an intermediate point $S_t = p_t S^\star_a+ (1-p_t)S^\star_0$, where $S^\star_0=0$ is the trivial solution and $S^\star_a$ is the power-law solution. 
The recursive  relation for $S$ reads:
\begin{equation}
	S_{t+1}(\mbr) =\frac{\sum_{\Delta \mbr_2 \in\{0,1\}^{n-1}} S_t(2\mbr-\Delta \mbr_2 )}{\sum_{\Delta \mbr_1, \Delta \mbr_2 \in\{0,1\}^{n-1}} C_t(\Delta \mbr_1-\Delta \mbr_2 )}=\frac{\sum_{\Delta \mbr_2 \in\{0,1\}^{n-1}} S_t(2\mbr-\Delta \mbr_2 )}{2^{n-2}\qty[1-f(S)]},
\end{equation}
where $f(S)$ is some linear combination of finite terms of $S(\mbr)$ whose values at the fixed points are $f(S_a^\star)=1-2^{n-a}$ and $f(S_0^\star)=0$. This leads to the recursive relation for $p_t$:
\begin{equation}
	p_{t+1} = \frac{2^{n-a}p_t}{1-p_t(1-2^{n-a})},
\end{equation}
which could be simplified to
\begin{equation}
	p_{t+1}^{-1}-1 = 2^{a-n}\qty(p_t^{-1}-1).
\end{equation}
Indeed, the critical exponent is $a_c=n$. The power-law fixed point is stable below $a_c$ and unstable above $a_c$. The dissipation scaling exponent is
\begin{equation}
	\lambda_{nd} = \frac{1}{n}\qty(1+\min(a,n)),
	\label{Eq:nDLambda}
\end{equation}
where $a$ characterizes the asymptotic decay of the fine-grained correlation.

Interestingly, the maximum possible scaling exponent $\max\lambda_{nd}=1+n^{-1}$ decreases monotonically with $n$. Its limit value in the infinite dimension limit $\lim\limits_{n\to\infty}\lambda_{nd} =1$ is exactly the exponent in the random hierarchical network (RHN), which could be considered the mean-field model of regular lattices~\cite{Yu2021}. This indicates the absence of any finite upper critical dimension, i.e., $n_c=\infty$. Therefore, the RHN model (random wiring) could never exactly represent regular lattice in finite dimensions.

\section{Discussion}
In this work, we developed a theoretical framework of State-Space Renormalization Group (SSRG) to study evolution of the correlation function between net probability fluxes in nonequilibrium reaction systems under coarse-graining of the network. The fixed points of the functional RG equation for the flux correlation function are solved exactly in hypercubic lattices to obtain a family of solutions with different power-law decay exponents. The stability and selection of these fixed point solutions are determined by studying the dynamics of the RG flow. 
These results can be directly used to explain the inverse power law dependence of energy dissipation rate on coarse-graining scales~\cite{Yu2021} and to determine the the corresponding dissipation scaling exponent $\lambda$.

The SSRG theory could be further developed to study many physical and biological systems driven out of equilibrium. A similar problem was recently studied with the Martin-Siggia-Rose field theory approach~\cite{Cocconi2021}, which is complementary to the SSRG theory. However, the SSRG approach is more amenable to generalization to other network structures, such as hexagonal lattices, scale free networks, or random hierarchical networks, which also exhibit dissipation scaling~\cite{Yu2021}. 
It would also be interesting to apply the SSRG theory to real biochemical networks and explore the biophysical significance of the dissipation scaling exponent $\lambda$.
More generally, the steady-state fluxes play an important role in characterizing various macroscopic physical observables of the NESS~\cite{Zia2007}, where the SSRG framework could prove useful.

The RG flow of the flux correlation function in 2D (or the radial system in higher dimensions) bears interesting similarity to the Kosterlitz-Thouless transition, where the exponent $a$ and the amplitude $(1-p_t^{-1})$ play the roles of effective temperature $K^{-1}$ and fugacity $y_0$, respectively. It might be useful to explore an exact mapping between the two systems to understand their similarity.
Surprisingly, the scaling behaviors in our system suggest that there is no finite upper critical dimension, which could be caused by the fact that there is no direct averaging effect among different transversal directions as suggested by the preservation of the angular distribution $f(\theta)$ under RG. Increasing the lattice dimension $n$ allows for more complicated angular dependence but does not change the radial equation, which determines $\lambda$. Thus, it should not bring any qualitative change of behavior with increasing $n$, suggesting no finite upper critical dimension.  
It may be interesting to explore its relation with other systems where the upper critical dimension is suggested to be infinite, such as the Kardar-Parisi-Zhang equation~\cite{Tu1994,Castellano1998,Canet2010}.

\begin{acknowledgments}
	The work by Y.~T. is supported in part by National Institutes of Health Grant No. R35GM131734. We thank Dr.~Dongliang Zhang for useful discussions. Q.~Y.~acknowledges stimulating discussions with Luca Cocconi.
\end{acknowledgments}

\bibliography{SSRG_refs}

\end{document}